\documentclass[conference,a4paper]{APSIPA2018}
\pdfoutput=1
\usepackage{amsmath}
\usepackage[numbers]{natbib}
\usepackage[T1, OT1]{fontenc}
\DeclareTextSymbolDefault{\dh}{T1}
\usepackage{hyperref}
\usepackage{color}
\usepackage{multirow}
\usepackage{listings}
\usepackage{algorithm}
\usepackage{algpseudocode}
\usepackage{graphicx}
\usepackage{threeparttable}

\begin{document}

\title{Tackling Android Stego Apps in the Wild}

\author{

\authorblockN{Wenhao Chen\authorrefmark{1}, Li Lin\authorrefmark{1}, Min Wu\authorrefmark{2}, and Jennifer Newman\authorrefmark{1}}

\authorblockA{
\authorrefmark{1}
Iowa State University, Ames, IA, USA\\
E-mail: \{wenhaoc, llin, jlnewman\}@iastate.edu
}

\authorblockA{
\authorrefmark{2}
University of Maryland, College Park, Maryland, MD, USA\\
E-mail: minwu@umd.edu
}
}

\maketitle


\begin{abstract}
Digital image forensics is a young but maturing field, encompassing key areas such as camera identification, detection of forged images, and steganalysis. However, large gaps exist between academic results and applications used by practicing forensic analysts. To move academic discoveries closer to real-world implementations, it is important to use data that represent ``in the wild" scenarios. For detection of stego images created from steganography apps, images generated from those apps are ideal to use. In this paper, we present our work to perform steg detection on images from mobile apps using two different approaches: ``signature'' detection, and machine learning methods.  A principal challenge of the ML task is to create a great many of stego images from different apps with certain embedding rates. One of our main contributions is a procedure for generating a large image database by using Android emulators and reverse engineering techniques. We develop algorithms and tools for signature detection on stego apps, and provide solutions to issues encountered when creating ML classifiers.
\end{abstract}


\section{Introduction}

Digital image forensics is a term used in academia to describe the study of digital images for camera identification, image forgery, and steganalysis. 
With the popularity of mobile devices, camera identification and forgery detection are attracting research for more practical scenarios for digital image forensics.
Important challenges remain, however, to detect steganography ``in the wild'' - produced by mobile apps - and where academic research may provide impetus for tool development.
In the penultimate case, evidence produced by a digital image forensic analysis for use in a court of law will be held to the Daubert standard \cite{stern2017statistical}, and assessed for scientifically-based reasoning and appropriate application to the facts.

Academic steganography and steganalysis techniques are very successful in the academic environment using sophisticated embedding and detection methods and data typically collected from digital still cameras~\cite{f5, nsf5, ste_adja_matrix, Wu, cc-jrm, ensemble}. While mobile phones appear as part of criminal investigations on a regular basis, surprisingly, the authors found only one published research on steganography detection where a mobile phone app was used to produce the stego images~\cite{chenifip2018}.  This paper presents our results of the first in-depth investigation into detection of stego images produced by apps on mobile phones.

In the academic community, a machine learning classifier is the first choice for steganalysis, and the algorithms are usually tested on a large database containing sufficient cover-stego pairs of images. However, as discussed in Section~\ref{section:dataset}, we show it is far from trivial to create appropriate training and testing data from these stego apps to use in machine learning (ML) classifiers. Moreover, unlike a published steganography algorithm,  Android developers prefer using more sophisticated techniques, including a password for encryption or a secret preprocessing package, or coding techniques such as obfuscation. The third challenge is the variety of devices and input images for the apps, and ignoring the impact of the image source can cause unacceptable errors in detection of stego images.
 
As Albert Einstein once said, in the middle of difficulty lies opportunity. For data collection, we develop an Android camera app~\cite{cameraw} that allows us to gather thousands of images by one device in just several hours. By using the most advanced tools from program analysis~\cite{soot, dexpler, apktool}, we make great efforts to reverse-engineer many Android stego apps. In analyzing the code written by developers of stego app programs, we determine that, with few exceptions, most algorithms used to hide the message are far from the advanced algorithms published in academic research papers. For example, some of the app embedding algorithms were based on simple least significant bit (LSB) changes placed in an lexicographical order in the image. Some apps provide little security, even if a complicated embedding method was applied, but strangely had a unique ``signature'' embedded, and make the stego image and its app easily identifiable as such. These are opportunities to take advantage of, and make steg detection easier.
 
Focusing on three phone models and seven Android stego apps, we present answers to our fundamental question using signature-based detection and machine learning classifiers. While admittedly these initial experiments are limited in scope, the experiments are soundly designed and provide the first such deep study published using images from mobile phone apps. We note that there is a large potential for security risks if these apps continue to be used, but lack effective detection. Table~\ref{t_apps_overview} shows 7 apps from the Google Play store, with a minimum of 1000 to 100,000 downloads each, as of May 2018. Along with iPhone stego apps, as well as apps not posted including the recently identified ``MuslimCrypt'' \cite{muslimcrypt}, there is evidence that steganography continues to be a model of digital communication. Understanding how to detect stego images from stego apps has potential to help accurately assess the rate that stego images occur ``in the wild.''

The remaining sections of the paper are as follows. In Section~\ref{section:background}, we discuss the background of steganography, how it is manifested in mobile apps, the general workflow of stego apps, and the critical challenges in creating thousands of images from mobile apps to use in steganalysis. Section~\ref{section:dataset} describes the important process of creating the image dataset for all our experiments, including the reverse-engineering procedure to generate the cover images corresponding to the app-generated stego images. In Section~\ref{section:signatures}, we present the results of the signature-based detection of stego images from mobile apps, followed by results of the machine learning methods in Section~\ref{S:2}. Finally, in Section~\ref{section:conclusion}, we summarize our work and look ahead to future challenges.
\section{Background}
\label{section:background}

\subsection{Steganography}
Steganography offers techniques to send a payload without revealing its presence in the medium of communication.
The focus of this paper is on digital images as the choice of medium for steganography. Unlike encryption of information, digital image steganography takes a payload, converts it to bits, and changes the bits of an image ever so slightly, so the changed image bits match the payload bits. This process is typically done in a way to avoid visual distortions of the image. In academic steganography, \textit{cover image} is the term for an image in its form just prior to embedding by a stego algorithm, and the term \textit{stego image} refers to an image output with hidden content. We call a \textit{message} the bit string corresponding to the user-input content desired to be communicated. It does not include passwords or other information generated by the app code, such as length of the payload or bit locations, for example. The term \textit{embed} is used to describe the algorithmic process of changing a cover image's bit values to represent a stego image's bit values. Typically, the change in bit values by the embedding algorithm results in change in color (or gray) intensity values by at most one. This leaves the overall visual content looking ``normal" to human vision. We use the term \textit{payload} to describe the combination of message, password, string length indicators and all other bit values that the app eventually embeds into the digital image. The \textit{payload size} is the number of bits needed to represent the payload.

\begin{figure}[t]
	\centering
	\vspace{-0.9cm}
	\includegraphics[width=0.40\textwidth]{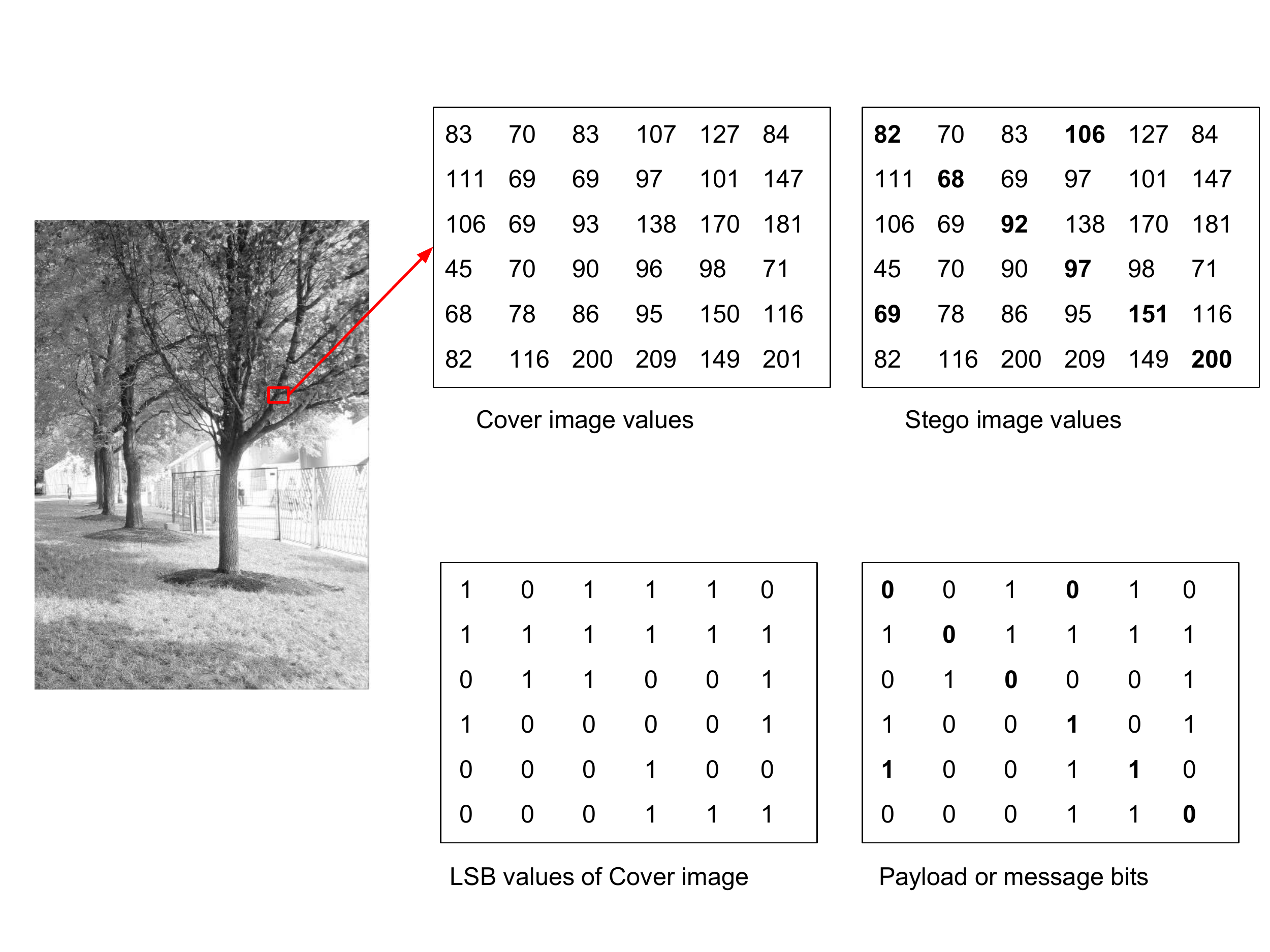}
	\vspace{-0.4cm}
	\caption{An example of LSB embedding in the spatial domain.}
	\vspace{-0.3cm}
	\label{lsb1}
\end{figure}

Many steganography algorithms exist to hide a payload in an image. One of the most common steganography algorithms uses the least significant bit values for embedding, and is adopted by many stego apps.
Figure \ref{lsb1} provides an example of one simple LSB-replacement embedding for a grayscale image in the spatial domain, in which all the changes are highlighted by bold numbers. In the stego images, the LSB  values of the cover images are replaced by the payload bits. 

To make the payload more secure when embedding, several additional steps can be followed. First, the payload bits themselves can be encrypted with a user-input, so that even if they are retrieved, the key is necessary for decrypting the payload. Second, the pixel locations where the payload bits are embedded can be selected in a random order, again using a key. As we will show in the following subsections, some stego apps use variations of the above methods to improve their level of security.

\begin{figure}[b]
\centering
\vspace{-0.3cm}
\includegraphics[width=0.35\textwidth]{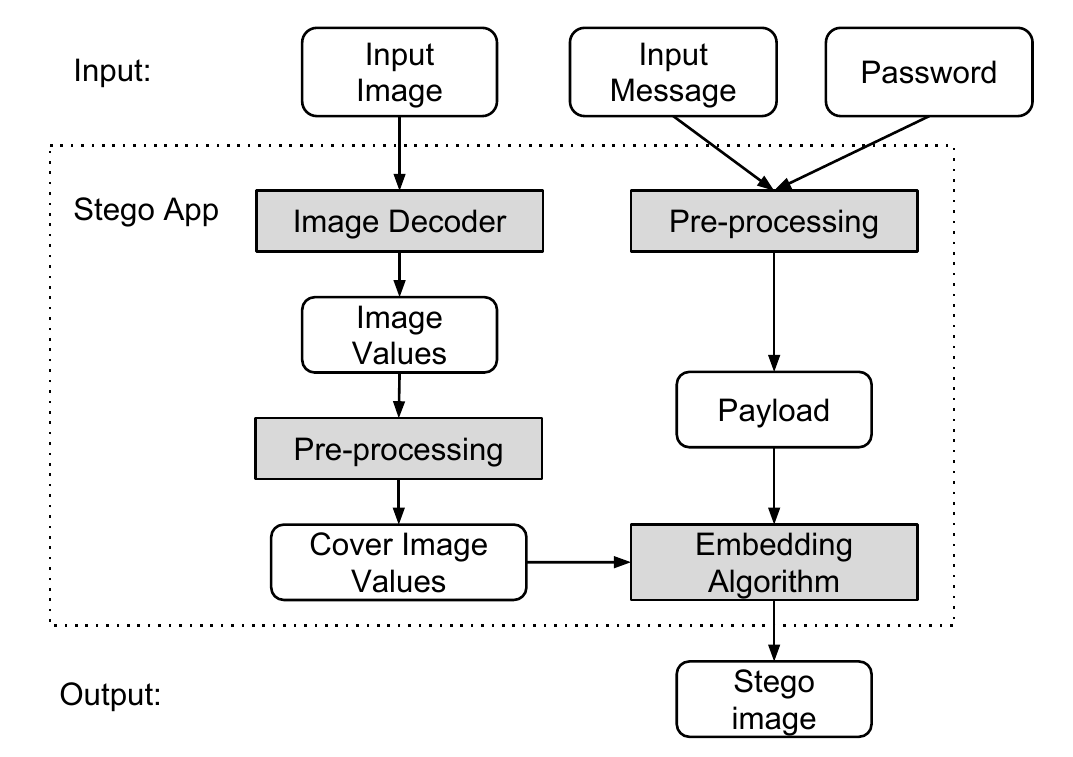}
\vspace{-0.2cm}
\caption{General Workflow of an Embedding Process.}
\label{f_workflow}
\end{figure}

\subsection{General Workflow of the Embedding Process in Stego Apps}

Although different stego apps may use different algorithms to create the stego images, they have many common features in their user interfaces. The user-input for stego apps include: (1) the input image, and (2) the message or file to be embedded, and optionally (3) the password. The output stego image is usually in PNG or JPEG format, depending on the image domain in which the payload is embedded.
Figure~\ref{f_workflow} shows the internal workflow of an embedding process in a stego app. Overall, an embedding process involves the following steps:
\begin{enumerate}
	\item Decode domain values of input image;
	\item Pre-process the domain values;
	\item Pre-process the message;
	\item Create and embed payload, and output stego image.
\end{enumerate}

First, the user-input image is decoded into a bitmap of pixel values, and then transformed into domain values. For spatial domain embedding, each domain value represents the RGB color and Alpha value of each pixel. For frequency domain embedding, the domain values represent the quantized DCT coefficients in the JPEG file. Additionally, the app may resize the cover image for different purposes such as reducing computational complexity, or increasing the cover image capacity.

Stego programs pre-process the message differently, such as encrypting the message before embedding to enhance security. Other stego programs attach signature strings or message length information so that the embedded message can be faithfully extracted by the receiver. A signature-based steg detection approach relies largely on the existence of the signature strings and length information. We provide more details on the signature-based detection approach in Section~\ref{section:signatures}.

\subsection{Steganalysis for Stego Images from Android Apps}
\label{section:background_steganalysis}
A steganalysis method has two steps: first, to discover if hidden payload is contained within the image, and, if so, extract and decrypt the hidden message. The vast majority of papers in the academic community has so far focused on classifying an image as cover (innocent) or stego (with hidden content). Machine learning has proved to be a successful method in classifying cover-stego pairs in academic settings. A typical ML framework for steganalysis includes a labeled image database, a corresponding feature space to represent the images, and a classification algorithm to separate the stego images from the clean images. The performance of a ML algorithm can be evaluated by the average misclassification rate for targeted images with a certain \textit{embedding rate}, where the \textit{embedding rate} is defined as:
$$ \frac{\text{\# of bits to represent the payload}}{\text{\# of the bits available to hide the payload (capacity) }}$$

One of the most popular image datasets in the academic community is BOSSbase \cite{Bas2011},  in which there are more than 10,000 cover images from seven digital still cameras. Using this dataset, steganalysts develop new algorithms to create stego images in a more secure way by sending the cover-stego pairs they create through advanced stego detectors. Most studies limit their cases to the balanced database scenario, in which the number of stego images is equal to the number of cover images, since under the assumption of balanced data, the average error rate of classification is sufficient to represent the performance of a steganalyzer.  That is, if we let $ P_{MD}$ denote the percentage of misdetections and $ P_{FA}$ the percentage of false alarms, then for a dataset constituting 50\% cover images and 50\% stego images, the average error rate $P_{E}$ for the detection is defined as:

\begin{equation}
P_{E}=\frac{1}{2} (P_{MD}+ P_{FA}).
\end{equation}

However, the scenario for detecting stego images created by mobile apps is different. First, unlike academic embedding algorithms or scripts which are capable of controlling the embedding rate by directly generating bit streams as payloads, a stego app encrypts real text together with a password (when available) into a bit stream as a payload for the target image. This means it is not trivial to generate a large amount of stego images at a specific embedding rate. Second, as we can see from Fig. \ref{f_workflow}, an input image provided by the user is processed before the embedding step in many of the stego apps, and therefore an input image is not necessarily the cover image in the traditional definition. Different stego apps apply different tools to process the input images into different image objects. In this paper, we view an image after the pre-processing procedure without any payload embedded as the \textit{cover} image. In this way, a cover image is a clean image produced by the same processing libraries applied to its stego pair, and the cover-stego pairs will have the same visual property that human eyes cannot tell the difference. Extracting a cover image as an intermediate output from a stego app is another challenge, and the details of creating cover images by reverse engineering are presented in the next section. 
\section{Generating Stego Image Dataset}
\label{section:dataset}

Since this is the first in-depth study on Android stego apps, a benchmarking database of images created by stego apps is essential. In this section, we describe the procedure of collecting original images, and the details of generating a large amount of stego images, including cover images and stego images at specific embedding rates from all Android stego apps that we have successfully reverse engineered.

\subsection{Collection of original images}

With the goal of studying Android apps, three mobile devices have been purchased,  which include a Google Pixel, a Samsung Galaxy S7, and a OnePlus 5. To better understand and control the quality of images captured by smartphone cameras, we develop an Android app named ``Cameraw"~\cite{cameraw} to collect original images. Cameraw allows us to take a group of ten pairs of images, including the DNG format and the JPEG format at same time, for each fixed scene with various exposure parameters in one click.
Table~\ref{t_data_originals} summarizes the original images captured for this study. A total of 421 different scenes of JPEG and DNG images were collected across all three devices. These original images are used to generate stego images from the 7 stego apps, which we introduce next.

\begin{table}[h]
\centering
\caption{Summary of Original Images collected from 3 smartphones}
\label{t_data_originals}
\begin{tabular}{l|r|r|r}
\hline
\multicolumn{1}{c|}{\textbf{Source Device}} & \multicolumn{1}{c|}{\textbf{\# Scenes}} & \multicolumn{1}{c|}{\textbf{JPEG}} & \multicolumn{1}{c}{\textbf{DNG}} \\ \hline
OnePlus 5 & 120 & 1200 & 1200 \\ 
Pixel1 & 187 & 1870 & 1870 \\ 
Samsung S7 & 114 & 1140 & 1140 \\ \hline
\textbf{Overall} & 421 & 4210 & 4210 \\ \hline
\end{tabular}
\vspace{-0.3cm}
\end{table}

\subsection{Generation of cover and stego images}
\subsubsection{Real-world Android Stego Apps}
To generate the stego images and their corresponding covers, we have chosen 7 of the most popular stego apps from the Google Play Store, as shown in Table~\ref{t_apps_overview}. We remark that the app \textit{Steganography\_M}~\cite{steganographyM} is actually named ``Steganography'' on Google Play Store. We append the letter M, which are the first letters of the author names, to distinguish the app with the many other stego apps also named ``Steganography'' on Google Play Store. As shown in the ``Output Format'' column, 2 of the 7 apps produce stego images in JPEG format while the other 5 produce PNGs. The two types of output format indicate different embedding domains: frequency domain embedding for JPEG and spatial domain embedding for PNG. The ``Open Source'' column shows that only 3 apps have their source code publicly available, which makes the app analysis process non-trivial.
We explain our process of reverse-engineering non-open source stego apps next.

\begin{table}[t]
\centering
\caption{Seven selected stego apps from Google Play Store}
\label{t_apps_overview}
\resizebox{0.5\textwidth}{!}{ 
\begin{tabular}{l|r|c|c}
\hline
\multicolumn{1}{c|}{\textbf{App Name}} & \multicolumn{1}{c|}{\textbf{\# Installs}} & \textbf{Output Format} & \textbf{Open Source} \\ \hline
PixelKnot & 100,000 - 500,000 & JPEG & yes \\
Steganography Master & 10,000 - 50,000 & PNG & no \\
Steganography\_M & 10,000 - 50,000 & PNG & no \\
Da Vinci Secret Image & 5,000 - 10,000 & PNG & no \\
PocketStego & 1,000 - 5,000 & PNG & no \\
MobiStego & 1,000 - 5,000 & PNG & yes \\
Passlok Privacy & 1,000 - 5,000 & JPEG & yes \\ \hline
\end{tabular}
}
\vspace{-0.3cm}
\end{table}

\subsubsection{Reverse engineering Android stego apps}
\label{section:reverseengineering}
As most of the stego apps are not open source, we utilize several Android program analysis tools to achieve reverse engineering. Given a stego app, we first use Apktool~\cite{apktool}, a reverse engineering tool for Android, to decode the program binaries and resource files from the app's APK file. The program binaries are decoded into an intermediate code format called Smali~\cite{smali}. The resources files are decoded into XML files that contain information about the app's graphical user interface (GUI). Next, we install and run the stego app on an Android device to test its user interface. We inspect the app's GUI structure while clicking through different screens, and use UIAutomator~\cite{uiautomator}, an Android GUI testing tool, to retrieve the resource IDs for different GUI widgets. Using the resource IDs, we then identify the GUI widget (usually a button named ``Embed'') that initiates the embedding procedure, and locate the corresponding event handler method in the Smali code. After the core embedding algorithm code is located, we manually inspect the code to understand the key characteristics of the embedding algorithm. Note that we use a different reverse engineering process for the app \textit{Passlok}, as \textit{Passlok} implements most of its functionality in JavaScript. We analyze Passlok's embedding algorithm from its publicly available JavaScript source code instead of the decoded Smali code.

\begin{table*}[t]
\centering
\caption{Characteristics of the embedding process in 7 stego apps}
\label{t_apps_details}
\resizebox{\textwidth}{!}{
\begin{threeparttable}
\begin{tabular}{l|c|c|c|c|c|c|c}
\hline
\multirow{2}{*}{\textbf{App Name}} & \multirow{2}{*}{\textbf{Embedding Domain}} & \multirow{2}{*}{\textbf{Image Resizing}} & \multicolumn{3}{c|}{\textbf{Payload Pre-processing}} & \multirow{2}{*}{\textbf{Embedding Path}} & \multirow{2}{*}{\textbf{Embedding Technique}} \\ \cline{4-6}
 &  &  & \textbf{Encryption} & \textbf{Signature String} & \textbf{Length Data} &  &  \\ \hline
PixelKnot & frequency & downsampling & yes & no & yes & pseudo-random & F5 \\
Steganography Master & spatial & no & no & yes & no & lexicographical & Base 10 LSD\tnote{a} \\
Steganography\_M & spatial & no & no & yes & yes & pseudo-random & LSB \\
DaVinci Secret Image & spatial & user-controlled & no & yes & yes & lexicographical & Alpha channel encoding \\
PocketStego & spatial & downsampling & no & yes & no & lexicographical & LSB \\
MobiStego & spatial & downsampling & yes & yes & no & regional lexicographical & RGB channels LS2B \\
Passlok Privacy & frequency & no & yes & yes & no & pseudo-random & ns-F5 \\ \hline
\end{tabular}
\begin{tablenotes}
\item[a] Least Significant Digit replacement in base 10, refers to replacing the ``ones'' digit in the cover image with one of the 3 base 10 digits of the message character, in each of the R-G-B planes.
\end{tablenotes}
\end{threeparttable}
}
\end{table*}

Our goal of reverse engineering stego apps, is to study the following characteristics of an embedding algorithm, so that we may batch-generate image data for our experiments:
\begin{itemize}
	\item \textbf{Embedding Domain}. The image domain in which the payload is embedded. It can be either frequency domain (JPEG) or spatial domain (PNG).
	\item \textbf{Image Resizing}. Indicates resizing of the cover image prior to embedding. Stego apps can downsize the input image to reduce computation time, and can upscale the input image to increase image capacity.
	\item \textbf{Payload Pre-processing}. The process that transforms the user input message into payload bits prior to embedding. For example, the input message can be encrypted, appended with a signature string or length data.
	\item \textbf{Embedding Path}. The order in which the domain values are visited to embed the payload. Some apps use simple lexicographical embedding paths, while others use pseudo-random embedding paths with the user password as a seed.
	\item \textbf{Embedding Technique}. How the payload is embedded into the domain values. Common embedding techniques are LSB embedding in the spatial domain and F5 embedding~\cite{f5} in the frequency domain. However, some of the apps we reverse-engineered have adopted their own unique embedding techniques.
\end{itemize}

Table~\ref{t_apps_details} shows the embedding characteristics of the 7 stego apps we reverse engineered. Note that ``user controlled'' in the column ``Image Resizing'' means the app lets the user decide the output image size. In the sub-columns of ``Payload Pre-processing,'' ``yes'' or ``no'' indicates the existence of payload processing steps.

The ``Embedding Domain'' column shows that two apps, \textit{PixelKnot} and \textit{Passlok}, embed payload into the DCT coefficients of the frequency domain, while the others embed payload in the spatial domain.

The ``Image Resizing'' column shows that 3 out of 7 apps do not resize the cover image, while 3 apps may downsample the input image. The app \textit{DaVinci Secret Image} allows the user to choose the image size from several options, including the option to maintain the original size.

The ``Payload Pre-processing'' column shows the three possible payload pre-processing options: encryption, signature string attachment, and length data attachment. Our investigation shows that prior to the embedding process, 3 apps perform encryption on the input message, 6 apps append signatures strings to the payload, and 3 apps append length data to the payload. While none of the apps perform all three payload pre-processing options, every app will attach at least either a signature string or a length information string to the payload. Such attachment to the payload is necessary for the app's extraction process, to identify the beginning and the end of the payload and correctly extract the message. However, it can leave patterns in the stego images that allow detection. We provide a detailed study on signature-based steg detection in Section~\ref{section:signatures}.

The ``Embedding Path'' column shows that 3 apps embed the payload along pseudo-random paths, while the others use fixed embedding paths. The pseudo-random embedding path is generated by a pseudo-random number generator using the user input password as seed. The embedding path can be recreated using the same seed. The lexicographical embedding path starts from the top left of the image, and proceeds row by row or column by column sequentially. The app \textit{MobiStego} uses a ``regional lexicographical'' order, where the cover image and payload are first split into multiple blocks, then a part of payload bits is embedded into each block lexicographically.

The ``Embedding Technique'' column shows the variety of embedding techniques in the chosen apps. The two frequency domain embedding apps \textit{PixelKnot} and \textit{Passlok} are both based on the academic embedding algorithm F5~\cite{f5}, where \textit{PixelKnot} removes the process of writing signature strings in the JPEG header, and Passlok uses a similar version of non-shrinkage F5~\cite{nsf5}. Out of the 5 spatial domain embedding apps, only \textit{Steganography\_M} and \textit{PocketStego} use the standard LSB Replacement, while \textit{DaVinci Secret Image} encodes the payload into alpha channel values. \textit{MobiStego} embeds 6 bits of payload into a single pixel by replacing the two least significant bits of all three RGB channels, and \textit{Steganography Master} embeds 8 bits of payload into one pixel by changing the decimal digit of each pixel's RGB value.

\subsubsection{Batch Image Generation through Instrumentation}
To achieve the goal of batch-generating stego images with fixed embedding rates while saving the intermediate cover images, we use binary instrumentation to change the apps' binaries to generate cover/stego pairs. The binary instrumentation is achieved by first decoding the APK file into Smali code, then modifying the Smali code to add functionalities, and finally compiling the modified Smali code back to an APK file using \textit{Apktool}~\cite{apktool}. Through instrumentation, we add two necessary functionalities to the stego apps: (1) saving the intermediate cover image along with its stego image pair, (2) automatically repeating the embedding process for all the input images.

\textbf{Saving Intermediate Cover Images}. As previously mentioned in Section~\ref{section:background_steganalysis}, statistical steganalysis benefits from having cover/stego pairs that went through the same image processing steps except for the embedding. However, real world stego apps, as shown in Table~\ref{t_apps_details}, often preprocess the input image, which makes the ideal ``cover'' image unavailable. The implementation of the functionality of saving the covers varies and depends on the specific stego app. In general, there are two ways to achieve this: (1) modify the app's embedding function so that it accepts ``empty payloads,'' in which case the produced stego image is equivalent to the cover image, or (2) add a new function to the stego app that can take the pre-embedding clean image data as input, and produce an image that has the same encoding or compression format as the stego image.
	
\textbf{Batch Cover/Stego Generation}. This functionality is necessary for processing the large amount of input images in our dataset in a timely manner. Each stego app has an added script module that recursively scans through folders of input images and calls the app's existing embedding functions to generate cover/stego pairs. Algorithm~\ref{alg_script} shows the pseudo code of the batch cover/stego generation script. The script has 3 inputs: a set of clean images, a dictionary pool for the input message, and a set of target embedding rates. For each input image, the script first pre-processes the image and saves the intermediate cover (Lines 2-3). The image capacity is then calculated (Line 4). Lines 6-13 generates messages with different target embedding rates and creates stegos. For each target embedding rate, we first calculate the length of the embedded payload \emph{lp}, then calculate the length of the input message \emph{lm} based on the knowledge acquired from reverse engineering the app. We then take a random segment of the dictionary with exact length \emph{lm}, and proceed to call the app's embedding function. Relevant stego information and statistics, including the message, password, embedding rate, change rate, etc., are also stored along with the stego image in the database.

\renewcommand{\algorithmicforall}{\textbf{for each}}

\begin{algorithm}[t]
\caption{Pseudo Code of Cover/Stego Generation Script}
\label{alg_script}
\hspace*{\algorithmicindent} \textbf{Input}: input\_images, dictionary, embedding\_rates
\begin{algorithmic}[1]
	\ForAll {$image$ in $input\_images$}
		\State $cover \gets \Call{Preprocess}{image}$
		\State $\Call{SaveImage}{cover}$
		\State $c \gets \Call{Capacity}{cover}$
		\ForAll {$rate$ in $embedding\_rates$}
			\State $lp \gets c \times rate$ \Comment embedded payload length
			\State $lm \gets \Call{Calculate}{lp}$ \Comment input message length
			\State $message \gets \Call{GetMessage}{dictionary, lm}$
			\State $password \gets \Call{GetPassword}{}$
			\State $payload \gets \Call{Preprocess}{message, password}$
			\State $stego \gets \Call{Embed}{cover, payload}$
			\State $\Call{SaveImage}{stego}$
			\State $\Call{SaveStegoInfo}{message, password}$
		\EndFor
	\EndFor
	\State \Return
\end{algorithmic}
\end{algorithm}

\begin{table}[b]
\centering
\vspace{-0.3cm}
\caption{Summary of Stego Images generated for 7 Stego Apps}
\label{t_data_stegos}
\resizebox{0.5\textwidth}{!}{ 
\begin{tabular}{l|c|c|r|r|r}
\hline
\multicolumn{1}{c|}{\textbf{Stego App}} & \multicolumn{1}{c|}{\textbf{Input Format}} & \multicolumn{1}{c|}{\textbf{Output Format}} & \multicolumn{1}{c|}{\textbf{\# Input}} & \multicolumn{1}{c|}{\textbf{\#Stegos}} & \multicolumn{1}{c}{\textbf{\# Covers}} \\ \hline
PixelKnot & JPEG+DNG & JPEG & 8420 & 42100 & 8420 \\
Steganography Master & PNG & PNG & 8420 & 42100 & 8420 \\
Steganography\_M & PNG & PNG & 8420 & 42100 & 8420 \\
DaVinci Secert Image & PNG & PNG & 8420 & 42100 & 8420 \\
PocketStego & PNG & PNG & 8420 & 42100 & 8420 \\
MobiStego & PNG & PNG & 8420 & 42100 & 8420 \\
Passlok & JPEG & JPEG & 4210 & 21050 & 4210 \\ \hline
\textbf{Overall} &   &  & 54730 & 273650 & 54730 \\ \hline
\end{tabular}
}
\end{table}

Table~\ref{t_data_stegos} shows the details of the stego images batch-generated from the 7 stego apps, using images in Table~\ref{t_data_originals} as input. For the two frequency domain apps, we use the original JPEG and DNG images to generate \textit{PixelKnot} stegos, and only the original JPEG images for \textit{Passlok} stegos as \textit{Passlok} cannot decode DNG images. For the other five spatial domain apps, we create PNG images that are center-cropped from the JPEG and DNG images, to reduce embedding time. For each input image, a total of 35 stego images are generated, including 5 stego images with different embedding rates from 7 different stego apps. The corresponding cover images for the stego images are also included in the dataset. In the next two sections, we present our study on signature-based steganalysis and machine learning steganalysis using the generated stego image dataset.

\section{Signature-based Steg Detection}
\label{section:signatures}

This section provides a study on signature-based steg detection. We first discuss the definition of signatures, then analyze the embedding signatures in 4 test apps, and present the signature-based detection results on stego images from the 4 apps.

\subsection{Embedding Signatures}

In the context of this paper, a \textit{signature} is a fixed pattern in the stego image that is unrelated to the cover image. In general, there are two types of embedding signatures: (1) fixed data written into fixed locations in the file headers, and (2) fixed data embedded into fixed locations in the image domain. As an example of signature type (1), the academic embedding algorithm F5~\cite{f5} writes comment messages into the JPEG file header, and uses a specific bit in the JFIF header~\cite{jfif} to indicate the existence of user comments. Signature type (2) appears more frequently in our surveyed apps such as \textit{DaVinci Secret Image} and \textit{Steganography Master}, where a fixed string pattern is embedded into fixed pixel locations.

The main reason that such signatures exist is to provide auxiliary information for the extraction of payload. The extraction process requires auxiliary information, such as length data or fixed signature strings, to identify the beginning and ending of the embedded payload. Subsequently, embedding signatures can be used to identify stego images and even potentially extract the payload.

As previously mentioned in Section~\ref{section:reverseengineering}, we use reverse engineering techniques to analyze the embedding algorithms from stego apps. As shown in Table~\ref{t_apps_details}, since the 3 apps \textit{PixelKnot}, \textit{Steganography\_M}, and \textit{Passlok} use pseudo-random embedding paths, their signature strings and length data are not in embedded in fixed locations. The other 4 apps \textit{Steganography Master}, \textit{DaVinci Secret Image}, \textit{PocketStego}, and \textit{MobiStego} each use a fixed lexicographical embedding path, which indicates a possible signature. Next, we analyze the embedding signatures of 4 stego apps and introduce our signature-based detection method.

\subsection{Signature-based Detection Approach}
\begin{figure}[t]
	\includegraphics[width=0.5\textwidth]{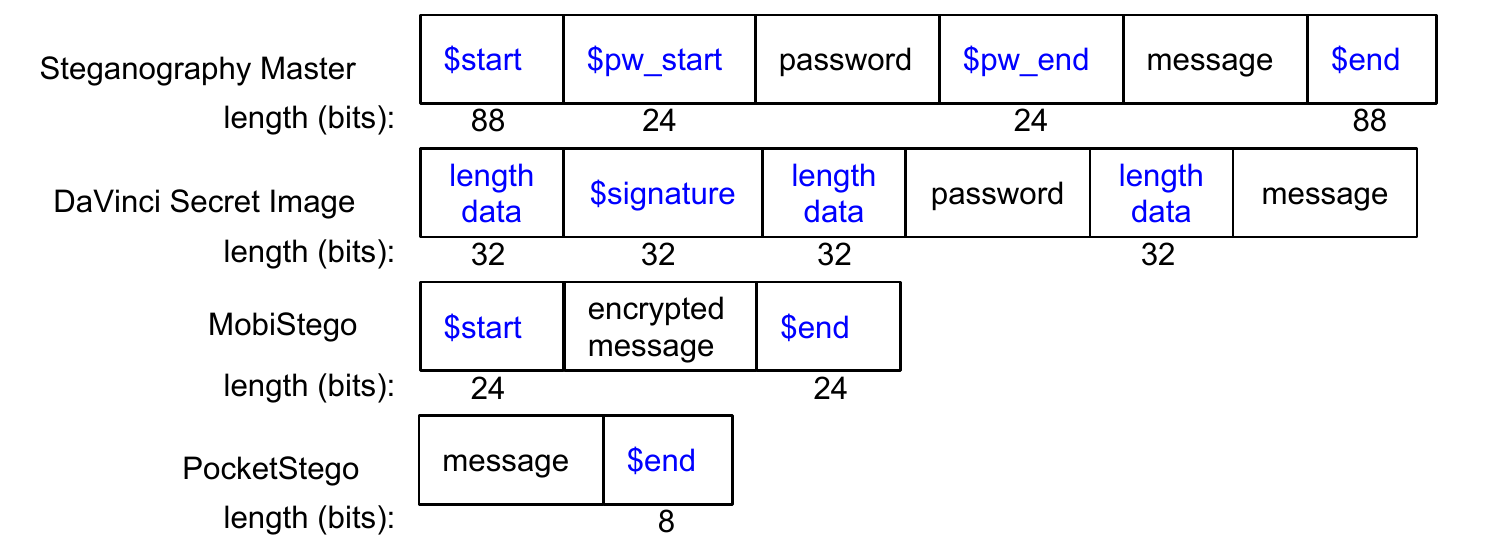}
	\vspace{-0.5cm}
	\caption{Format of Processed Payload in 4 Stego Apps.}
	\vspace{-0.5cm}
\label{f_payload_formats}
\end{figure}

The formats of embedded payload of the 4 stego apps are shown in Fig~\ref{f_payload_formats}. Each app has a different way of converting the user input message and/or password into the embedded payload. Attached to the messages and passwords are two types of data: (1) signature strings which are represented with label ``\$'', (2) length data, which is only used by \textit{DaVinci}. The app \textit{Steganography Master} joins the password and input message in plaintext and surrounds them with two pairs of fixed signature strings. The embedded payload in \textit{DaVinci} consists of three segments: the signature string, the password in plaintext, and the message in plaintext. Each segment is also prefixed with a length data whose value is the number of bits of the segment. The app \textit{MobiStego} first encrypts the input message using the password, then surrounds the encrypted message with a pair of signature strings. The app \textit{PocketStego} only appends a short 8-bit signature string at the end of the plaintext message.

To utilize the signature strings for steg detection, we also nmust know the apps' embedding paths and embedding techniques. The embedding path determines the pixel locations in which the signature string is embedded, and the embedding technique decides how to embed and extract the payload bit(s) from each pixel data. Given that all 4 stego apps use lexicographical embedding paths, and that 3 of them have fixed-length signature data at the start of the payload, we can identify the pixel locations that might contain these unique signature data.

The embedding techniques of the 4 stego apps have been reverse engineered, as shown in Table~\ref{t_apps_details}. \textit{Steganography Master} first turns each 8 bits of payload into a decimal number ranging from 0 to 255. The 3 digits of the decimal number then replace the least significant base 10 digits of a pixel's R, G, B decimal values, respectively. \textit{DaVinci} embeds 1 payload bit per pixel by setting the pixel's alpha value to 254 if the payload bit is 0, or to 255 if the payload bit is 1. \textit{MobiStego} embeds 6 payload bits per pixel by replacing the least significant two bits of all three RGB values. \textit{PocketStego} uses the standard LSB Replacement embedding where each payload bit overwrites a pixel's LSB in the Blue channel only.

With the knowledge of the signature strings, embedding paths, and embedding techniques from the 4 stego apps, we implement 4 stego detection functions. Each stego detection function corresponds to one of the stego apps. Each detection function takes a test image as input, and outputs a decision on whether the test image is a stego image produced by the stego app. The decision is made by extracting the embedded bits based on the stego app's embedding pattern, and check whether the extracted payload matches with the correct payload format. Next, we present our experimental results on signature-based steg detection. 

\subsection{Experimental Result}
The image dataset for this experiment contains 202,080 images including 168,400 stego images and 33,680 cover images from the 4 stego apps: \textit{Steganography Master}, \textit{DaVinci Secret Image}, \textit{MobiStego}, and \textit{PocketStego}.
The test results are shown in Table~\ref{t_result_both}. For each detector, the test data is grouped into two categories: (1) stego images generated from this stego app (labeled as \textit{SM}, \textit{DV}, \textit{MS}, \textit{PS} for abbreviation), and (2) all other images, including cover images and stego images from the other 3 stego apps. The detection results for the two groups of data are shown separately for each stego detector. As the results show, the 3 stego detectors for \textit{Steganography Master}, \textit{DaVinci Secret Image}, and \textit{MobiStego} correctly identify all stego images generated from their corresponding apps, while correctly distinguishing these stego images from cover images and other stego images. While the \textit{PocketStego} detector has correctly identified all the \textit{PocketStego} stego images, it also mis-identifies the majority of the other images.
 
\begin{table}[h]
\centering
\caption{Results of Signature-based Steg Detection}
\label{t_result_both}
\resizebox{0.5\textwidth}{!}{ 
\begin{tabular}{l|l|r|r}
\hline
\multicolumn{1}{c|}{\textbf{Stego App}} & \multicolumn{1}{c|}{\textbf{Test Images}} & \multicolumn{1}{c|}{\textbf{Image Count}} & \multicolumn{1}{c}{\textbf{Accuracy}} \\ \hline
\multirow{2}{*}{Steganography Master} & SM Stego Images & 42,100 & 100\% \\
 & Other Images & 159,980 & 100\% \\ \hline
\multirow{2}{*}{DaVinci Secret Image} & DV Stego Images & 42,100 & 100\% \\
 & Other Images & 159,980 & 100\% \\ \hline
\multirow{2}{*}{MobiStego} & MS Stego Images & 42,100 & 100\% \\
 & Other Images & 159,980 & 100\% \\ \hline
\multirow{2}{*}{PocketStego} & PS Stego Images & 42,100 & 100\% \\
 & Other Images & 159,980 & 0.23\% \\ \hline
\end{tabular}
}
\end{table}

The prefect results for \textit{Steganography Master}, \textit{DaVinci Secret Image}, and \textit{MobiStego} detectors are as expected, as these apps have very distinctive signature strings in their payload. For example, as shown in Fig~\ref{f_payload_formats}, \textit{Steganography Master} has two fixed signature strings (112 bits in total), \textit{DaVinci Secret Image} has 64 bits of distinct signature strings, and \textit{MobiStego} has 24 bits of distinct signature strings, at the beginning of their payloads. On the other hand, \textit{PocketStego} has only one 8-bit signature string at the end of the payload, without a fixed location. This ``weak'' signature can be found in not only the stego images from PocketStego, but also randomly occurs in 99.77\% of other images as well, resulting in very poor accuracy.

Our test results demonstrate that it is possible to detect real world stego images based on app embedding signatures. The advantage of signature-based steg detection is that, by looking for known signatures in a test image, it can reliably identify stego images, and perhaps even extract the embedded payload. However, signature-based steg detection relies on the uniqueness of signature strings. Longer signature strings with more distinct patterns yield better detection, while shorter signature strings potentially lead to mis-identification. Another limitation of signature-based detection is that some apps use (unknown) random embedding paths. Furthermore, the extraction of the signature is also not a trivial process. Manual inspection of the app binaries is not an optimal approach, especially when the app has anti-analysis features such as obfuscation or native code. Our future work on signature-based detection is to automate the process of extracting signatures from stego apps using program analysis methods.

\section{Detecting Stego Apps by Machine Learning Algorithms}
\label{S:2}

Although we achieve almost perfect results for detecting stego apps that involve signatures, many stego apps do not write any signature to the images they create. For those apps without signatures, we use machine learning methods. In this section, two Android apps, PixelKnot and Steganography, are selected for our studies, using two well-known methods: the F5 algorithm, and spatial LSB embedding. To the best of our knowledge, this is the first time a ML detection algorithm is applied to identify stego images generated by mobile stego apps. We implement the CC-JRM \cite{cc-jrm} for feature extraction on JPEG images and SRM \cite{srm} for feature extraction on PNG images. The FLD ensemble classifer \cite{ensemble} performs do the classification.

\subsection{Case Study - PixelKnot}

PixelKnot \cite{pixelknot} is an Android  app that uses a modified version of the embedding algorithm F5 \cite{f5} and it outputs a stego image in JPEG format. Before embedding, PixelKnot downsamples the input image if its size exceeds 1280*1280. The message is encrypted using one part of the password, while the other part of the password is used as seed for the F5 algorithm to generate a pseudo-random embedding path.

\subsubsection{Experiments and Results}

The goal of our first experiment is to study if the academic ML methods can be applied to detect the stego images created by PixelKnot. For every selected original image (DNG or JPEG), PixelKnot loads the standard Picasso package to downsample this larger image into a smaller bitmap object, which can be viewed as a 8-bit image in the spatial domain\footnote{We use grayscale images in our experiments.}. Cover and stego images with different embedding rates are generated from all original images  by reverse engineering as discussed in the previous section. To evaluate the performance of the classic ML method, we implement CC-JRM and FLD ensemble classifiers for feature extraction and classification, respectively.

To give the first impression of ML detector for stego apps, for every device, we randomly select 850 original JPEG images as the input, and create the cover-stego pairs from those originals. We use 500 for training, and 350 for testing. The results are presented in Table \ref{pixelknot1}.

\begin{table}[h]
	
	\caption{Detecting 10\% stego images created by PixelKnot with 10\% payload, where the input images are JPEG format.}
	\centering
	
	\resizebox{90mm}{!}{
\begin{tabular}{l|c|r}
\hline
\textbf{\textbf{Data Source}} & \textbf{\textbf{Original Image Size}} & \multicolumn{1}{l}{\textbf{Avg. Error Rate}} \\ \hline
Google Pixel & 4048 $ \times $ 3036 & \textbf{1.0\%} \\
Samsung Galaxy S7 & 4032 $ \times $ 3024 & \textbf{7.6\%} \\
OnePlus 5 & 4608 $ \times $ 3456 & \textbf{0.3\%} \\ \hline
Mixture of above three devices & Flexible & \textbf{3.2\%} \\ \hline
\end{tabular}
	}
	
	\label{pixelknot1}
\end{table}

As we can see from the Table \ref{pixelknot1},  when the image data sources are fixed, the results are quite encouraging. Even in the case when we mix the images from all three devices, the average error rate is just about 3\%. Table \ref{pixelknot1} shows that with the knowledge of the suspect image data source, for a fixed embedding rate, an academic ML based detection method works well in detecting the stego images generated from PixelKnot.

We have to point out that the results in Table \ref{pixelknot1} are based on the assumption that we have the knowledge of the source devices of the suspected images. However, this is not typical in a real-world scenario. The scenario when the source of the target image is not in the training database is called the \textit{cover-source-mismatch problem}  in steganalysis \cite{csmismatch}. Table \ref{pixelknot2} lists the results in the case when the sources of test images are not involved in the training database.

\begin{table}[h]
	
	\caption{Detecting stego images created by PixelKnot with 10\% payload, in the cover-source-mismatch case.}
	\centering
	
	\resizebox{90mm}{!}{
		\begin{tabular}{l|c|r}
\hline
\textbf{\textbf{Test Data Source}} & \textbf{\textbf{Training Data Source}} & \multicolumn{1}{l}{\textbf{Avg. Error Rate}} \\ \hline
Google Pixel & Samsung Galaxy s7 \&  OnePlus 5 & \textbf{1.3\%} \\
Samsung Galaxy s7 & Google Pixel \&  OnePlus 5 & \textbf{40.0\%} \\
OnePlus 5 & Google Pixel \&  Samsung Galaxy s7 & \textbf{35.7\%} \\ \hline
\end{tabular}
	}
	
	\label{pixelknot2}
\end{table}

In Table \ref{pixelknot2},  the average error rates are not at the same level for the three cover-source-mismatch cases. Although the error in testing images from Google Pixel is almost as low as the error in Table \ref{pixelknot1}, the error rates of detecting stego images from the other two devices when they are out of the training datasets, are much  greater than those  in Table \ref{pixelknot1}. Our preliminary results show that it is not desirable to use only one or two devices to build a classifier for blind detection on multiple devices.  Thus, knowing the source of the target images in detecting stego apps will significantly reduce the error rate for a ML based analyzer.

\begin{table}[h]
	\caption{Average error rates for detecting stego images created by PixelKnot with different payload sizes.}
	\centering
	\resizebox{90mm}{!}{
		\begin{tabular}{l|r|r|r|r}
\hline
\multicolumn{1}{r|}{\textbf{\quad Training Set:}} & \multicolumn{1}{l|}{\multirow{2}{*}{\textbf{5\% Stego}}} & \multicolumn{1}{l|}{\multirow{2}{*}{\textbf{10\% Stego}}} & \multicolumn{1}{l|}{\multirow{2}{*}{\textbf{15 \% Stego}}} & \multicolumn{1}{l}{\multirow{2}{*}{\textbf{20\% Stego}}} \\
\textbf{Test Set:} & \multicolumn{1}{l|}{} & \multicolumn{1}{l|}{} & \multicolumn{1}{l|}{} & \multicolumn{1}{l}{} \\ \hline
\textbf{5\% Stego} & 7.9\% & 41.4\% & 47.9\% & 49.6\% \\
\textbf{10\% Stego} & 4.7\% & 1.0\%\ & 12.9\% & 39.3\% \\
\textbf{15\% Stego} & 3.9\% & 0.9\% & 0.4\%\ & 2.86\% \\
\textbf{20\% Stego} & 4.4\% & 0.7\% & 0.4\% & 0.2\% \\ \hline
\end{tabular}
	}
	
	\label{pixelknot3}
\end{table}

\begin{figure}[t]
	
	\centering	
	\includegraphics[width=65mm]{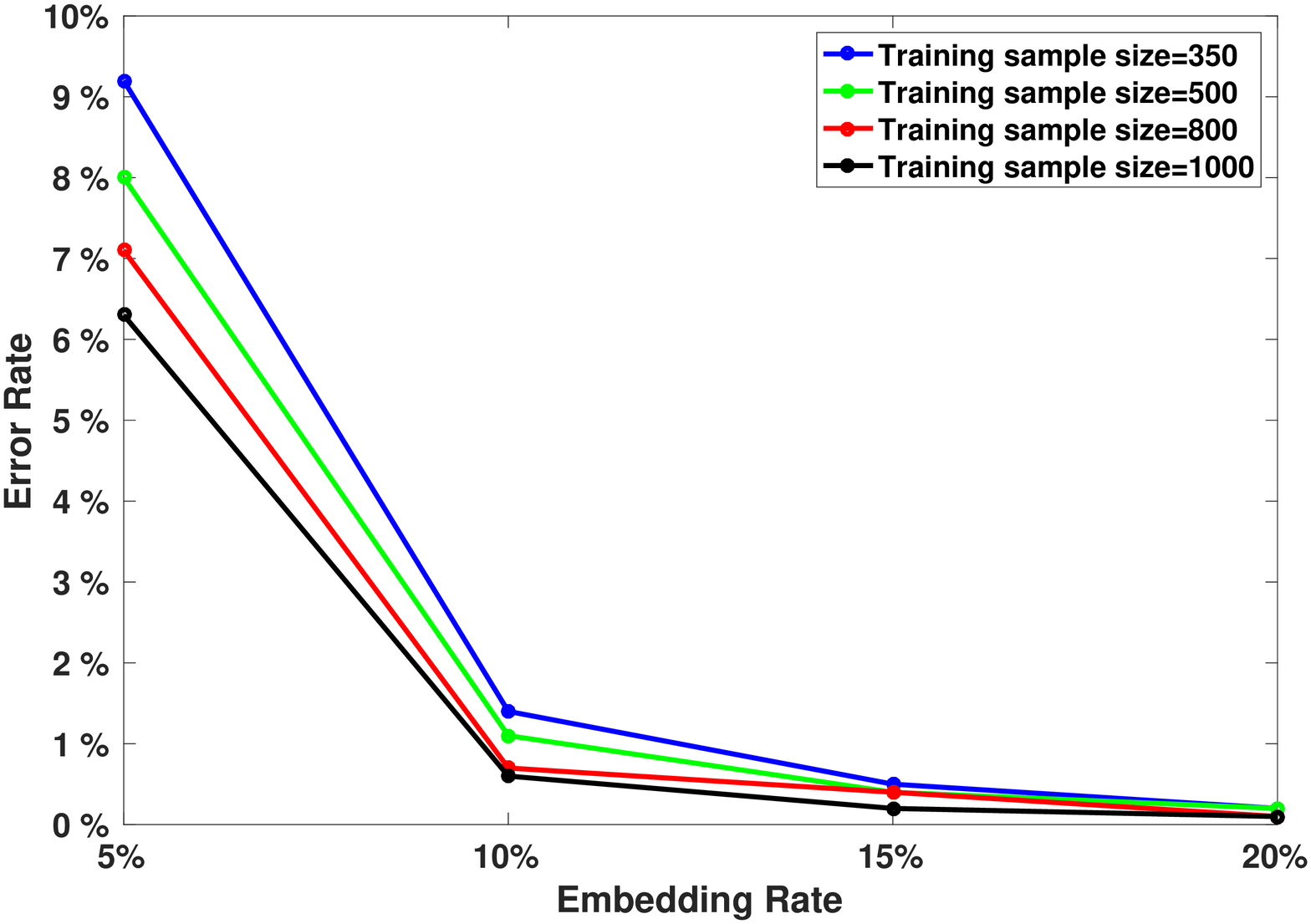}
	\vspace{-0.1cm}
	\caption{Detecting stego images created by PixelKnot, where the original images are JPEG images collected from Google Pixel\label{Pixelknot_plot}.}
	
	\centering	
	\includegraphics[width=65mm]{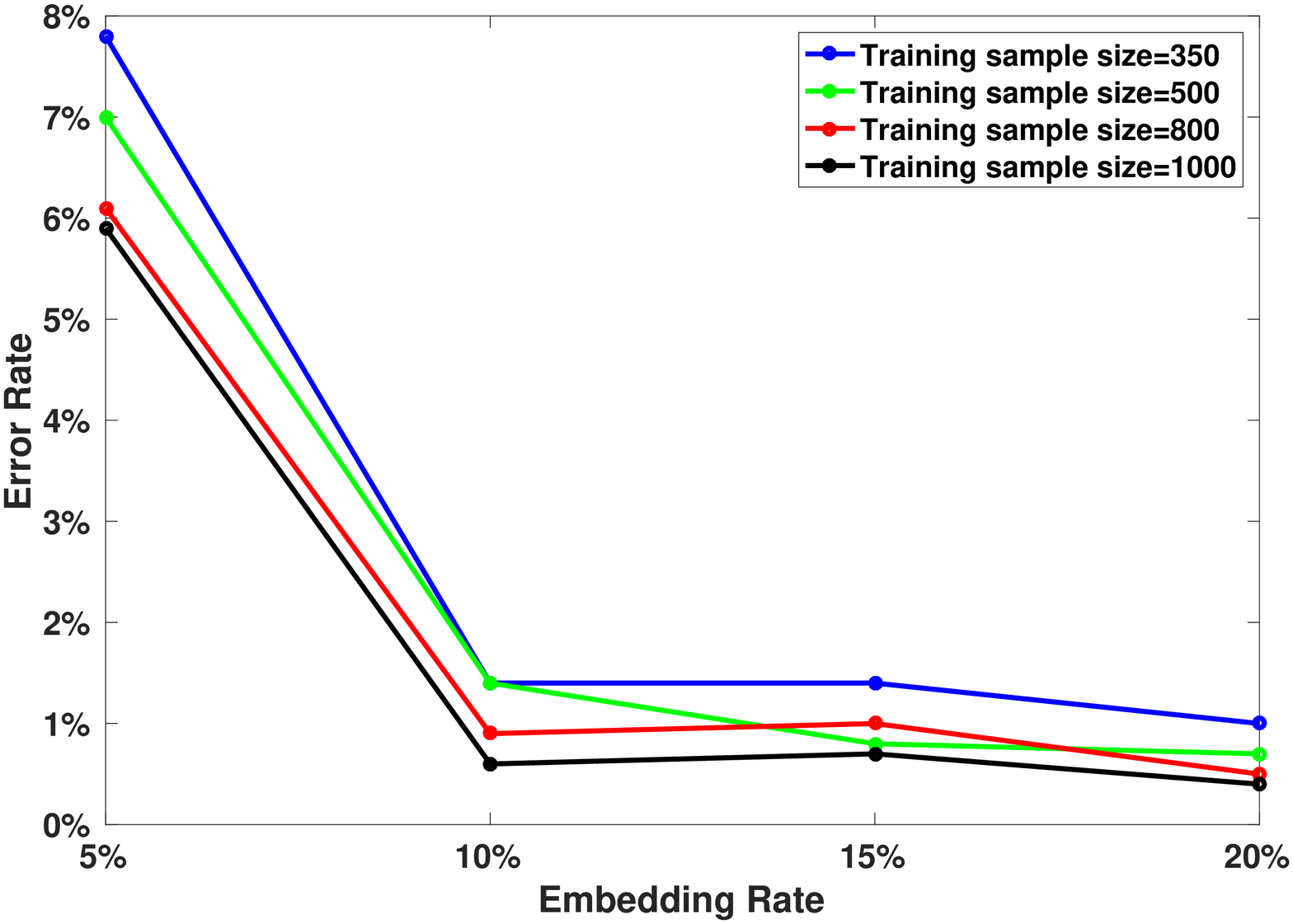}
	\caption{Detecting stego images created by PixelKnot, where the original images are DNG images collected from Google Pixel\label{Pixelknot_plot2}.}
	\vspace{-0.5cm}

\end{figure}

In the previous experiment, the embedding rate is fixed at 10\% for all stego images we created. To study the possibility of detecting stego images with flexible embedding rates, we use the previous input images collected from Google Pixel to generate three more sets of stego images with three different embedding rates: 5\%, 15\% and 20\%. For each subset, we build a stego classifier for a fixed embedding rate. The results are presented in Table \ref{pixelknot3}.

In Table \ref{pixelknot3}, for the same test data, the lowest error rate always occurs in the diagonal entries, for which the training stego images have the same embedding rate as the images for testing. Another interesting phenomenon is that the many error values located above the diagonal are extremely high, while error values below the diagonal look acceptable. Table \ref{pixelknot3} gives us an impression that it may be possible to apply well-trained stego classifiers based on 5\% embedding rate to suspect images having unknown payloads. 
We have to admit that this conclusion is limited to the case when all images are from only one mobile phone, which is the Google Pixel, and it is left to future research for the verification by larger-scale experiments for a variety of sources.

The performance of a ML based steg detector can depend on the training sample size and the embedding rate for the target data. To test this, we randomly select 350 cover-stego pairs from the Google Pixel's images to create the target data with varied embedding rate, and then randomly select another groups of images with different sample size to train, repeating ten times. The result is plotted in Fig. \ref{Pixelknot_plot} and Fig. \ref{Pixelknot_plot2}. It is clear in Fig. \ref{Pixelknot_plot} and Fig. \ref{Pixelknot_plot2} that with the increase in embedding rate, the average error rates by training with all four sets decrease. Also, adding more images into the training dataset will slightly reduce the error, especially in the case when the embedding rate is very low. Last but not the least, since Pixelknot preprocesses and downsamples the input images, the choices we made for the input images (JPEG or DNG)  did not cause significantly different results between these two experiments.

%
\subsection{Case Study - Steganography\_M}
To study how well a ML detection method works in the case when stego images are created by spatial domain embedding algorithms, we run our second case study for the app Steganography\_M~\cite{steganographyM}. Steganography\_M is an Android stego app that uses spatial domain embedding with pseudo-random embedding paths and implements an embedding algorithm very similar to the standard LSB spatial embedding.  From the clues we have found during the program analysis, we know that the cover image is not resized before embedding, and the pseudo-random embedding path is generated from the user password. 

\subsubsection{Experiments and Results}
Since there is no preprocessing by Steganography\_M, the noise levels of input images can affect the detection results significantly. For that reason,  two original image formats, JPEG and DNG are both considered in this case study. However, with limited time and resources, we use cropped grayscale PNG images with dimension 512 $ \times$ 512 from original images as the input images. One benefit of such a choice is that, feature extraction for smaller images is much more efficient. Another important reason is that using processed PNG images for the input can be viewed as clean cover images, since they are not compressed or downsampled by the app. With the help of Google Android emulator, we create thousands of stego images with varied embedding rates for all original images (JPEG and DNG) collected by the app Cameraw installed on all three devices.

In this experiment, SRM and ensemble classifiers are applied for feature extraction and classification, respectively. For each fixed embedding rate, we randomly select a sample of images for training and then test the classifier on another sample. Table \ref{sm_jpeg}  provides the first result when JPEG images are used to generate the inputs. As we can see from Table \ref{sm_jpeg}, when original images are in JPEG format, the ML based steg detectors work so well that even for a very low embedding rate ($<$ 5\%), the error rates are never above 3\% for all devices. 
\begin{table}[h]
	\caption{Detecting stego images created by Steganography\_M ( original images are JPEG, training sample size = 500, test sample size =350).}
	\centering
\begin{tabular}{c|c|r}
\hline
\textbf{\textbf{Data Source}} & \textbf{\textbf{Embedding Rate}} & \multicolumn{1}{c}{\textbf{Avg. Error Rate}} \\ \hline
\multirow{3}{*}{Google Pixel} & 3\% & \textbf{0.4\%} \\
 & 5\% & \textbf{0.0\%} \\
 & 8\% & \textbf{0.0\%} \\ \hline
\multirow{3}{*}{Samsung Galaxy S7} & 3\% & \textbf{1.3\%} \\
 & 5\% & \textbf{0.9\%} \\
 & 8\% & \textbf{0.4\%} \\ \hline
\multirow{3}{*}{OnePlus 5} & 3\% & \textbf{2.7\%} \\
 & 5\% & \textbf{1.4\%} \\
 & 8\% & \textbf{1.0\%} \\ \hline
\end{tabular}
	\label{sm_jpeg}
\end{table}

As we did for Pixelknot, the results of using fixed embedded data to detect stegos with different embedding rates is summarize in Table \ref{sm_dif}. In this experiment, only JPEG images from Google Pixel are used to generate input images, and for every classifier, 500 random pairs of cover-stego images are used for training, 350 for testing. The results in Table \ref{sm_dif} are very similar to what we concluded for the app Pixelknot. 

\begin{figure}[t]
	
	\centering	
	\includegraphics[width=60mm]{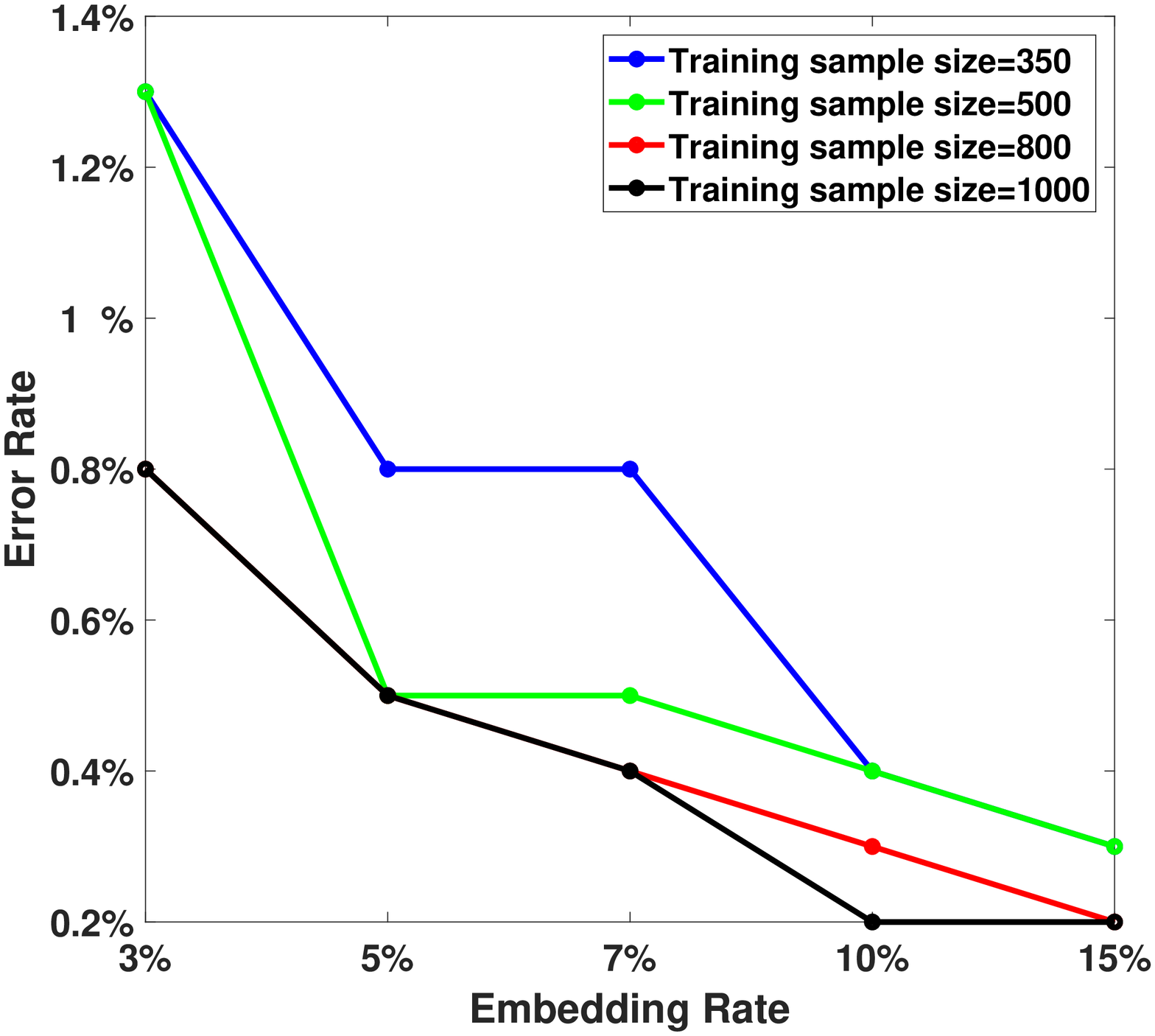}

	\caption{Detecting stego images created by Steganography\_M, where the original images are JPEG images collected from Google Pixel\label{sm_plot}.}
	\centering	
	\includegraphics[width=60mm]{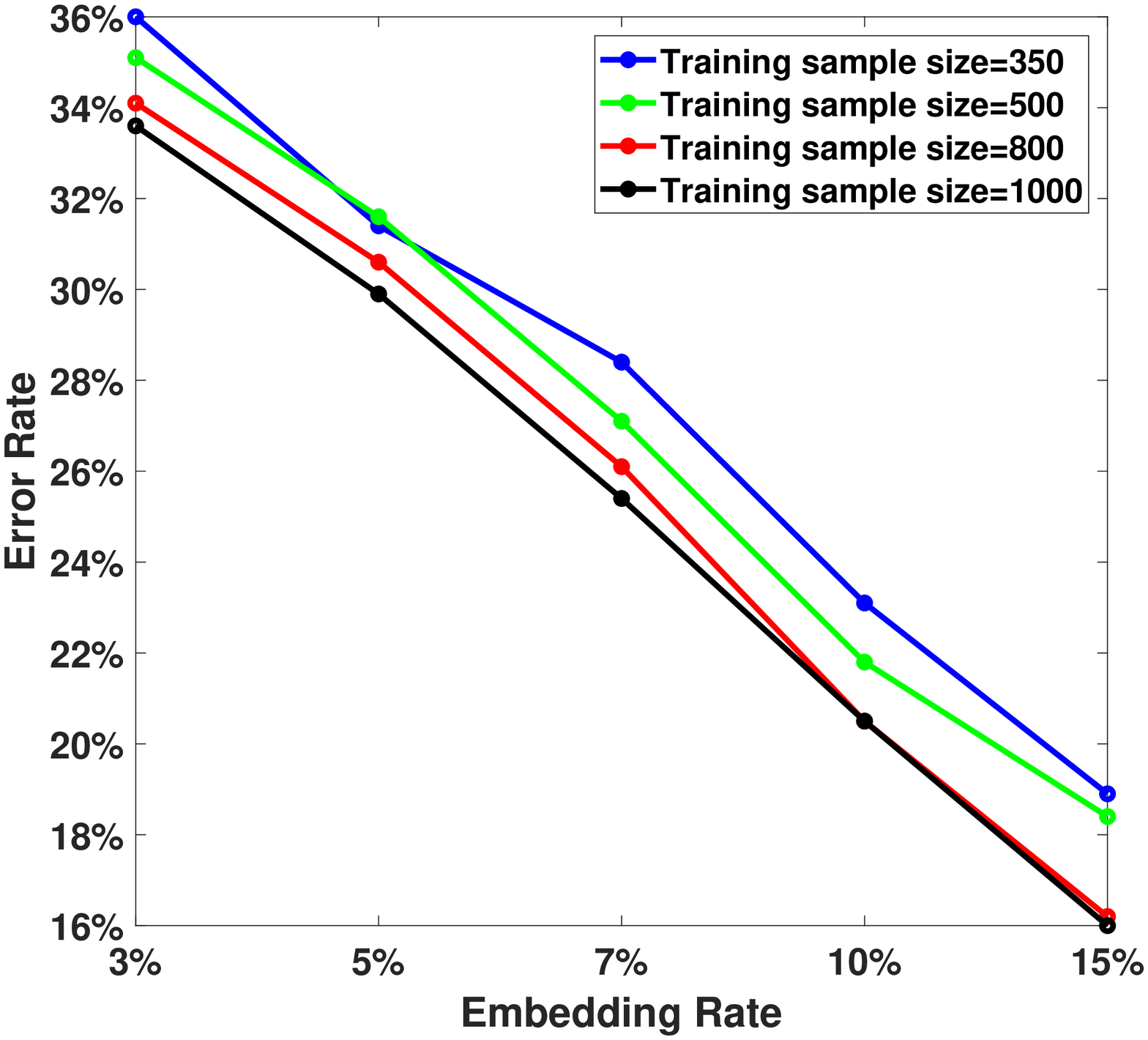}

	\caption{Detecting stego images created by Steganography\_M, where the original images are DNG images collected from Google Pixel\label{sm_plot2}.}
\end{figure}
 
\begin{table}[h]
	\caption{Average error rates for detecting stego images created by Steganography\_M with different payload sizes.}
	\centering
	\resizebox{90mm}{!}{
		\begin{tabular}{l|r|r|r|r}
\hline
\multicolumn{1}{r|}{\textbf{\quad Training Set:}} & \multicolumn{1}{l|}{\multirow{2}{*}{\textbf{2\% Stego}}} & \multicolumn{1}{l|}{\multirow{2}{*}{\textbf{3\% Stego}}} & \multicolumn{1}{l|}{\multirow{2}{*}{\textbf{5 \% Stego}}} & \multicolumn{1}{l}{\multirow{2}{*}{\textbf{7\% Stego}}} \\
\textbf{Test Set:} & \multicolumn{1}{l|}{} & \multicolumn{1}{l|}{} & \multicolumn{1}{l|}{} & \multicolumn{1}{l}{} \\ \hline
\textbf{2\% Stego} & 2.9\% & 28.4\% & 48.0\% & 49.6\% \\
\textbf{3\% Stego} & 2.4\% & 0.4\%\ & 0.6\% & 28.4\% \\
\textbf{5\% Stego} & 2.7\% & 0.6\% & 0.0\% & 0.1\% \\
\textbf{7\% Stego} & 2.6\% & 0.4\% & 0.0\% & 0.0\% \\ \hline
\end{tabular}
	}
	\vspace{-0.4cm}
	
	\label{sm_dif}
\end{table}

However, as we mentioned above, there is no preprocessing procedure for the input images by \textit{Steganography\_M}. As a result, compared to using JPEG as the data source, the noisy DNG images significantly increases the detection errors. To illustrate this phenomenon, with the device fixed to Google Pixel, we randomly select 350 JPEG images and 350 DNG images to create cover-stego pairs as the test dataset at different embedding rates, repeat ten times, and for each time and each embedding rate, we create the corrsponding training datasets with four different sample sizes. The results are present in Fig \ref{sm_plot}. and Fig  \ref{sm_plot2}. Figures \ref{sm_plot} and \ref{sm_plot2} show that using the DNG images to generate the input images for \textit{Steganography\_M} produces higher error from a ML classifier. Moreover, even with 1000 cover-stego pairs for training and a high embedding rate of 15\%, the error of misclassification is at a minimum of 16\% for DNG images from one phone.

Based on the above experiments, we conclude that, comprehending the algorithms of stego apps is very important for detecting stego images created by apps. Therefore, the concept of ``cover-source" should also be applied to the detection of stego images from stego apps. So far, we focus on two Android apps, one for frequencey domain embedding, and one for spatial domain embedding. In both cases, we assume that we have the knowledge of the app that created stego images. But this is not common in practice. In most cases, we may have very little information about the stego apps. To explore the possibility of using one well-known app to detect the stego images created by a different app, our future work is to use more apps, such as \textit{Passlok}, to study blind detection.
\section{Conclusion}
\label{section:conclusion}
In this paper, we analyze seven Android apps that implement steganography algorithms. A major contribution of this paper is our procedure to analyze the code in the app by applying reverse-engineering techniques to the binary code.
We use instrumentation techniques to perform the non-trivial task to batch-generate cover-stego image pairs for machine learning steganalysis.
Thus, with appropriate numbers of images, we create machine learning classifiers and perform successful steganalysis on stego images created from two mobile apps. We also present a detailed analysis of four stego apps that contain a signature, and perform steg detection on these images. While this is currently done in a manual process, our future efforts will investigate methods to automate the program analysis of app code. Another challenge is to go beyond the task of identifying stego or innocent images: the extraction of hidden contents.  A program analysis approach can also be useful to solve this problem. Another future challenge we plan to implement are other learning paradigms, including deep learning.

\section{Acknowledgement}
\begin{small}
This work was partially funded by the Center for Statistics and Applications in Forensic Evidence (CSAFE) through Cooperative Agreement \#70NANB15H176 between NIST and Iowa State University, which includes activities carried out at Carnegie Mellon University, University of California Irvine, and University of Virginia.
\end{small}

\bibliographystyle{abbrv}
\bibliography{References}

\end{document}